\documentclass[twocolumn,tighten]{aastex62}
\def\kms{{km s$^{-1}$}}

\shorttitle{Small-Scale Coherence-Conformity}
\shortauthors{Lee et al.}
\def\simlt{\lower.5ex\hbox{$\; \buildrel < \over \sim \;$}}
\def\simgt{\lower.5ex\hbox{$\; \buildrel > \over \sim \;$}}

\begin{document}
\title{Small-Scale Dynamical Coherence Accompanied with Galaxy Conformity}

\author{Joon Hyeop Lee}
\email{jhl@kasi.re.kr}
\author{Mina Pak}
\author{Hye-Ran Lee}
\affil{Korea Astronomy and Space Science Institute, Daejeon 34055, Republic of Korea}
\affil{University of Science and Technology, Daejeon 34113, Republic of Korea}

\begin{abstract}
The discovery of the coherence between galaxy rotation and neighbor motion in 1-Mpc scales has been reported recently. Following up the discovery, we investigate whether the neighbors in such dynamical coherence also present galaxy conformity, using the Calar Alto Legacy Integral Field Area Survey (CALIFA) data and the NASA-Sloan Atlas (NSA) catalog. We measure the correlation coefficient of $g-r$ colors between the CALIFA galaxies and their neighbors, as a quantitative indicator of galaxy conformity. The neighbors are divided into coherently moving and anti-coherently moving ones, the correlation coefficients from which are compared with each other, in various bins of relative luminosity and projected distance. In most cases, the CALIFA galaxies and their neighbors show positive correlation coefficients in $g-r$ color, even for the anti-coherent neighbors. However, we find statistically significant ($2.6\sigma$) difference between coherent and anti-coherent neighbors, when the neighbor galaxies are bright ($\Delta M_r\le-1.0$) and close ($D\le400$ kpc). That is, when they are bright and close to the CALIFA galaxies, the coherently moving neighbors show stronger conformity with the CALIFA galaxies than the anti-coherently moving neighbors. This result supports that the small-scale dynamical coherence may originate from galaxy interactions as galaxy conformity is supposed to do, which agrees with the conclusion of the previous study.
\end{abstract}


\section{INTRODUCTION}\label{intro}

Since the integral field spectroscopy (IFS) started to spatially resolve the spectroscopic information of a galaxy, our understanding of galaxy kinematics has been significantly improved. Now, it is generally believed that most galaxies, not only late-type galaxies but also early-type galaxies, are rotating \citep[e.g.,][]{cap06,ems07}, and galaxy rotation has risen as an important clue to trace the formation history of a galaxy. Galaxy rotation is known to largely depend on the interactions with close neighbors \citep[e.g.,][]{kra15,kra18,oh16,lee18,wea18} as well as large-scale environments \citep[e.g.,][]{tem13,hir17,kim18,jeo19,lee19b}. Both of the speed and the direction of galaxy rotation are of our great interest, the origins of which are being actively investigated in recent studies.

On the origin of the direction of galaxy rotation, one of the recent and notable findings is that the rotational direction of a galaxy tends to be coherent with the average line-of-sight motion of its neighbor galaxies in 1-Mpc scales \citep{lee19a}. Such coherence appears to be particularly strong for the galaxy's outskirt rotation (at $R_e<R\leq 2R_e$; where $R_e$ is the effective radius) and when the rotating galaxy is faint while its neighbor galaxy is bright. From this observational evidence, \citet{lee19a} concluded that a fly-by interaction with a massive neighbor may significantly change the rotational direction of a galaxy, particularly at its outskirt.

If the origin of the small-scale dynamical coherence is recent interactions (possibly in a few Gyr) between galaxies as argued in \citet{lee19a}, one may expect that such interactions influence not only galaxy kinematics but also the photometric properties of the galaxies, because interacting galaxies or close pairs tend to share their properties such as morphology or stellar populations \citep[galaxy conformity;][]{wei06,ann08,par08a}. That is, galaxy interactions that may induce dynamical coherence, may also cause the conformity between galaxies. Even though the conditions for dynamical coherence and galaxy conformity may not exactly coincide with each other, we may find some specific conditions in which they happen at the same time.

Thus, to verify the previous finding of \citet{lee19a} and to better understand the details of what happens to galaxies that have recently experienced interactions, we investigate whether the dynamical coherence in 1-Mpc scales is accompanied with galaxy conformity, which has been never studied before. Our key question is: `does galaxy conformity appear to be different between the neighbors with coherent and anti-coherent motions?’
This paper is outlined as follows: Section~\ref{data} describes the data set and our method to examine the relationship between dynamical coherence and galaxy conformity in a small scale. The results are presented in Section~\ref{result}, and their implication is discussed in Section~\ref{discuss}. Throughout this paper, we adopt the cosmological parameters: $h=0.7$, $\Omega_{\Lambda}=0.7$, and $\Omega_{M}=0.3$.

\begin{figure}[t]
\centering
\plotone{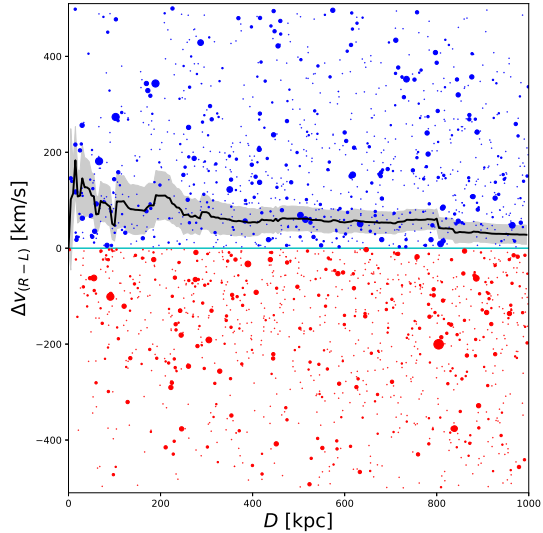}
\caption{Relative velocity ($\Delta v_{\textrm{\tiny (R-L)}}$) versus projected distance ($D$) distribution of the neighbors around the CALIFA galaxies, showing the key result of \citet{lee19a}. A positive/negative $\Delta v_{\textrm{\tiny (R-L)}}$ indicates a coherent/anti-coherent motion of each neighbor to the outskirt rotation of a given CALIFA galaxy (blue/red filled circles). The symbol size represents the relative luminosity of the neighbor to a given CALIFA galaxy (a larger symbol for smaller $\Delta M_r\equiv M_r(\textrm{neighbor}) - M_r(\textrm{CALIFA})$). The cumulative luminosity-weighted mean velocity profile (black line) and its random-spin-axis uncertainty (shaded area) are overlaid.\label{coher}}
\end{figure}

\begin{figure*}[t]
\centering
\plotone{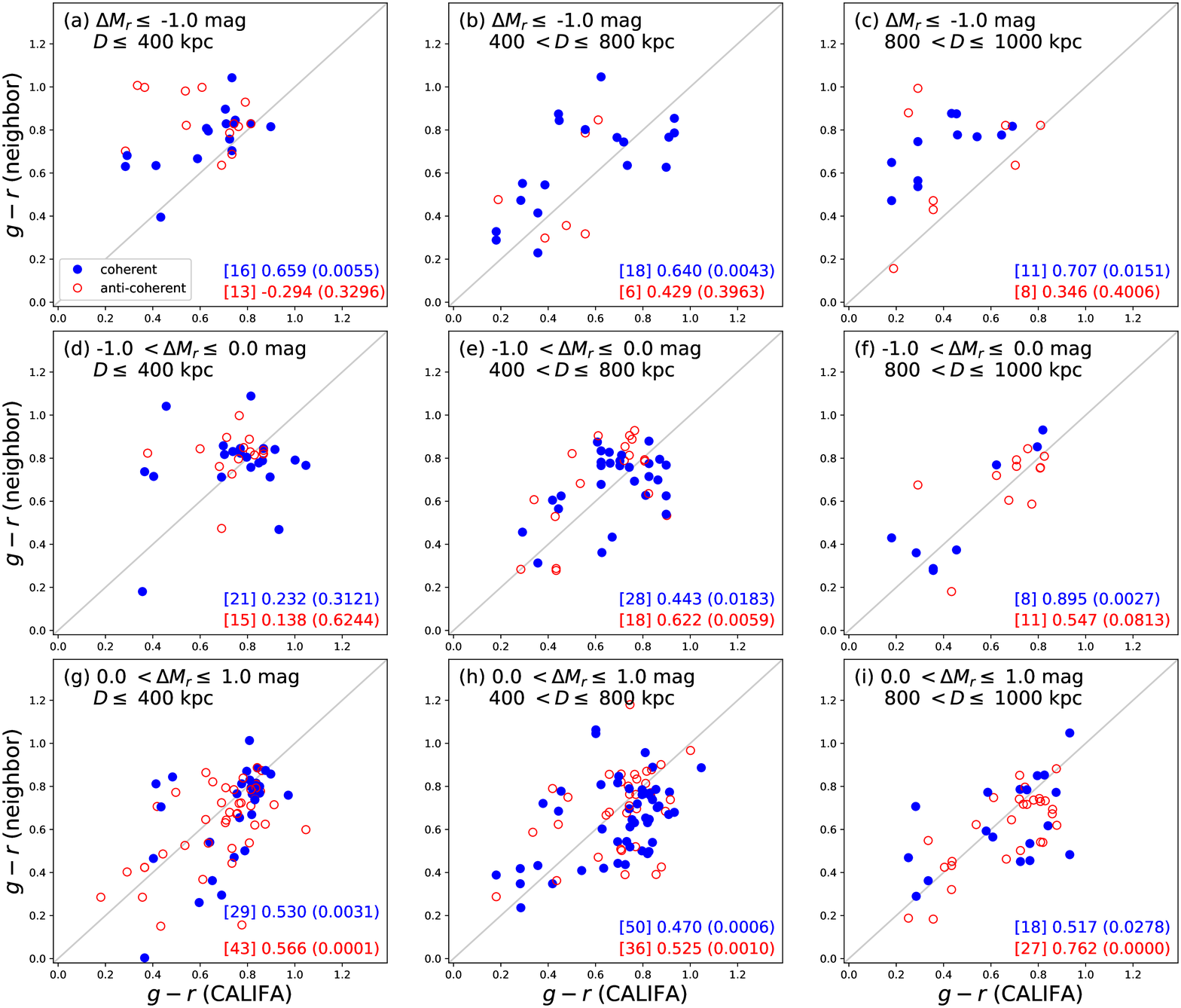}
\caption{Comparison of $g-r$ color between the CALIFA galaxies and their neighbors, which are divided by their relative luminosities and projected distances. In each panel, coherent (blue filled circles) and anti-coherent (red open circles) neighbors are marked by different symbols. The numbers at the bottom-right corner in each panel show: [the number of CALIFA - neighbor pairs] Pearson correlation coefficient ($p$-value). \label{col9}}
\end{figure*}

\begin{figure*}[t]
\centering
\plotone{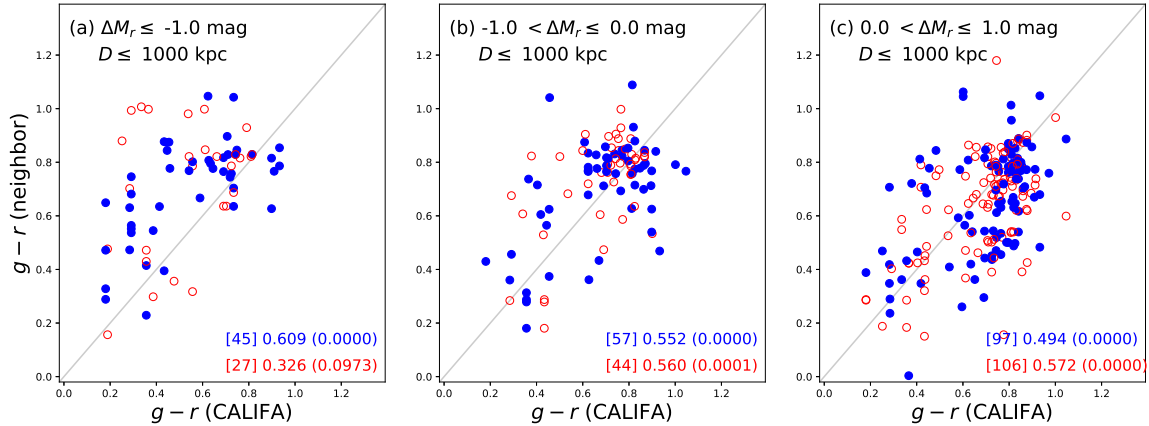}
\caption{Comparison of $g-r$ color between the CALIFA galaxies and their neighbors within 1 Mpc, which are divided by their relative luminosities (i.e., the merged version of Figure~\ref{col9} along $D$ out to 1 Mpc). \label{col3}}
\end{figure*}

\begin{figure*}[t]
\centering
\plotone{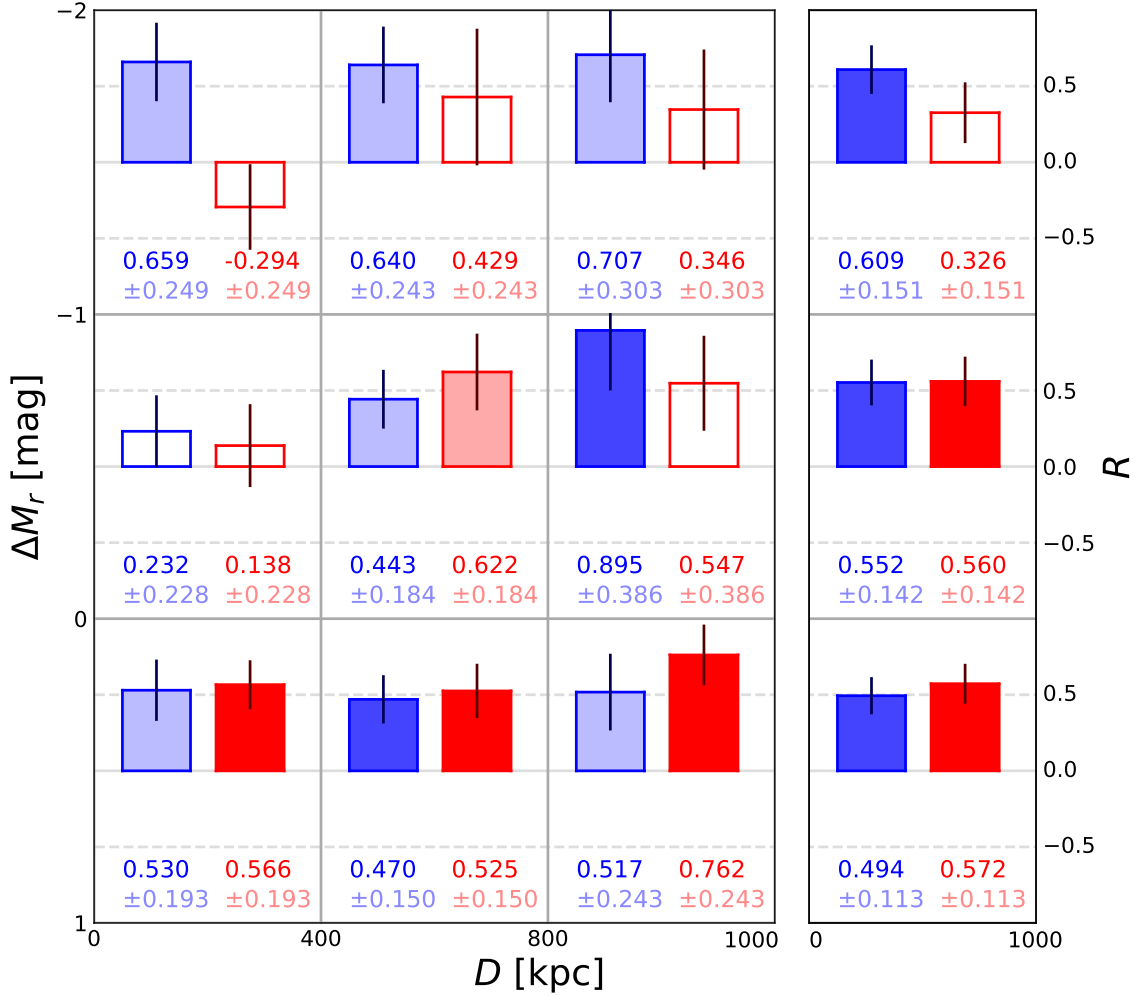}
\caption{Comparison between coherent and anti-coherent neighbors, of conformity in $g-r$ color. Blue/red wide-bars show the Pearson correlation coefficients ($R$) between the CALIFA galaxies and their coherent/anti-coherent neighbors. The brightness of each wide-bar color reflects the $p$-value: $p\le0.003$ (dark), $0.003<p\le 0.050$ (light), and $p>0.050$ (white). The bootstrap uncertainty of each $R$ is denoted by an errorbar on each wide-bar.
The correlation coefficient $\pm$ bootstrap uncertainty is also given in number for each case.\label{rcomp}}
\end{figure*}

\begin{figure*}[t]
\centering
\plotone{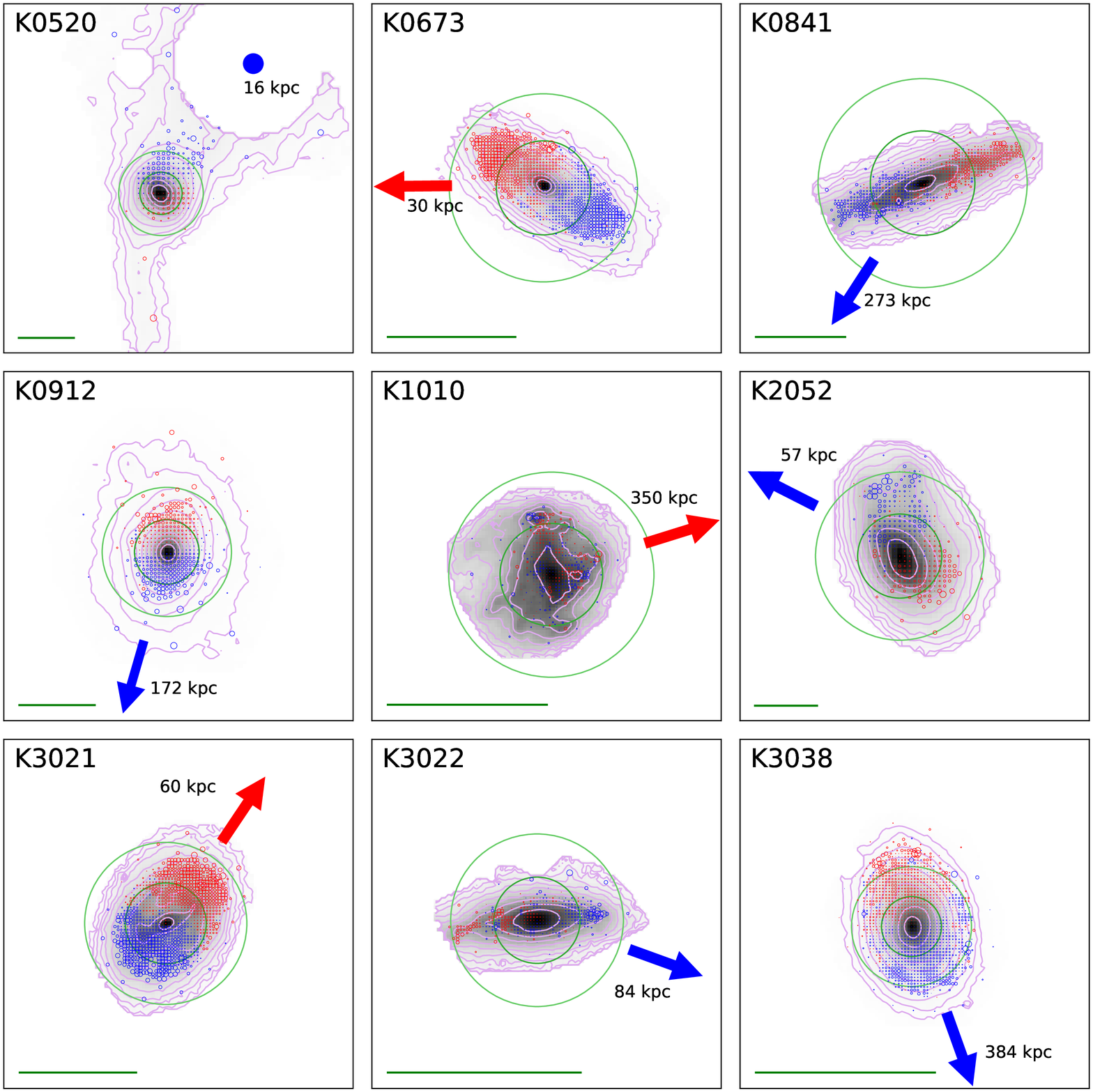}
\caption{Several examples of the CALIFA galaxies that show both of the dynamical coherence and the photometric conformity with their bright neighbors ($\Delta M_r \le-1.0$) at $D\le400$ kpc. The inverted-black-and-white image of each CALIFA galaxy is overlaid by its surface brightness contours (purple), which is also overplotted by resolved velocity ingredients (red circles when receding, while blue circles when approaching). Two green circles represent the effective radius and its twice. The arrow directs toward the bright neighbor, and its projected distance is denoted beside it. The color of the arrow indicates the relative line-of-sight velocity of the bright neighbor: red (receding) or blue (approaching). K0520 is the only case that the bright neighbor is closer than 30 kpc (and in the field of the CALIFA observation), in which ongoing interactions are clearly visible. The green bar at the lower-left corner in each panel shows the 5 kpc distance scale. \label{atlas}}
\end{figure*}

\section{DATA AND METHOD}\label{data}

As done in \citet{lee19a}, we use the PyCASSO database\footnote{http://pycasso.ufsc.br or http://pycasso.iaa.es/} \citep{dea17}, which is a value-added data set from the Calar Alto Legacy Integral Field Area Survey \citep[CALIFA;][]{san12,san16,wal14} data. Among the 445 CALIFA galaxies at $z\lesssim0.03$ from the PyCASSO database, the angular momentum vectors at galaxy outskirts ($R_e<R\le 2R_e$) were derived for 392 targets. Then, the line-of-sight motions of their neighbor galaxies were estimated, using the NASA-Sloan Atlas (NSA) catalog\footnote{http://www.nsatlas.org}. The NSA catalog was created by Michael Blanton, by combining the Sloan Digital Sky Survey \citep[SDSS;][]{yor00}, NASA Extragalactic Database (NED)\footnote{https://ned.ipac.caltech.edu/}, Six-degree Field Galaxy Redshift Survey \citep[6dFGS;][]{jon09}, Two-degree Field Galaxy Redshift Survey \citep[2dFGRS;][]{col01}, CfA Redshift Survey \citep[ZCAT;][]{huc83}, Arecibo Legacy Fast ALFA Survey \citep[ALFALFA;][]{gio05}, and the Galaxy Evolution Explorer \citep[GALEX;][]{mar03} survey data.

We select the neighbor galaxies of a given CALIFA galaxy with the X-cut configuration described in \citet{lee19a}: only the neighbors between $45^{\circ}$ and $135^{\circ}$ or between $225^{\circ}$ and $315^{\circ}$ from the CALIFA angular momentum vector direction are selected, in the line-of-sight velocity cut of $\pm500$ {\kms}. Although the sample size of neighbor galaxies is reduced by half in the X-cut, this selection significantly improves the reliability in distinguishing between coherent and anti-coherent motions of neighbors. For example, suppose that there are  two neighbor galaxies with the same properties but the position angles from the angular momentum vector of a given CALIFA galaxy to be $5^{\circ}$ and $355^{\circ}$, respectively. In this case, the actual difference in their position angles is only $10^{\circ}$, and thus they will make almost the same impacts on the CALIFA galaxy. However, without applying the X-cut, the two galaxies are regarded as one coherent neighbor and one anti-coherent neighbor respectively, and thus their influence on the CALIFA galaxy will be counted oppositely to each other, which is not so reasonable. Moreover, when it is considered that we limited the uncertainty in measuring the angular momentum vector direction up to $45^{\circ}$ \citep[Figure~5 in][]{lee19a}, the X-cut is necessary to guarantee the reliability in the estimation of dynamical coherence.

Figure~\ref{coher} revisits the key result in \citet{lee19a}. The relative velocity, $\Delta v_{\textrm{\tiny (R-L)}}$ indicates the difference in line-of-sight velocity between a CALIFA galaxy and its neighbor, which is positive/negative when the moving direction of the neighbor is coherent/anti-coherent with the rotational direction of the CALIFA galaxy. Overall, $\Delta v_{\textrm{\tiny (R-L)}}$ is widely distributed over the coherent ($>0$) and anti-coherent ($<0$) domains.  However, the cumulative mean velocity profile weighted by relative luminosities of neighbors keeps positive values within 1 Mpc, and its statistical significance at $D=800$ kpc is as large as $3.0\sigma$. This plot shows that a larger number of neighbors tend to move coherently to the rotational directions of the CALIFA galaxies. However, there are also many anti-coherently moving neighbors, the number of which is never negligible.

Note that multiple neighbors can be chosen for a given CALIFA galaxy in our selection, for which we count each CALIFA-neighbor relationship independently. Thus, it is possible that a CALIFA galaxy has coherently-moving neighbors and anti-coherently-moving neighbors at the same time, but more massive neighbors tend to be more coherent to the rotation of the CALIFA galaxy, according to \citet{lee19a}.

In this paper, we investigate if there is any meaningful difference in the conformity with the CALIFA galaxies, between the coherent ($\Delta v_{\textrm{\tiny (R-L)}}>0$) and anti-coherent ($\Delta v_{\textrm{\tiny (R-L)}}<0$) neighbors. As a simple measure of conformity, we use the Pearson correlation coefficient between the $g-r$ colors of the CALIFA galaxies and their neighbors: a positively strong correlation indicates good conformity. Since the dynamical coherence appears to significantly depend on relative luminosity ($\Delta M_r\equiv M_r(\textrm{neighbor}) - M_r(\textrm{CALIFA})$) and projected distance to a given CALIFA galaxy ($D$) as shown in \citet{lee19a}, we compare the correlations in various bins of $\Delta M_r$ and $D$.

\section{RESULTS}\label{result}

\begin{deluxetable*}{ccr@{ $\pm$ }lcr@{ $\pm$ }lcr@{ $\pm$ }lc}
\tablenum{1} \tablecolumns{11} \tablecaption{Conformity Difference Between Coherent and Anti-Coherent Neighbors} \tablewidth{0pt}
\tablehead{  Luminosity & Distance [kpc] & \multicolumn{2}{c}{$R_C$} & $p_C$ & \multicolumn{2}{c}{$R_{AC}$} & $p_{AC}$ & \multicolumn{2}{c}{$R_C - R_{AC}$} & $\sigma$ }
\startdata
 & $D\le400$ & $0.659$ & $0.249$ & 0.0055 & $-0.294$ & $0.272$ & 0.3296 & $0.953$ & $0.369$ &  2.58 \\
$\Delta M_r\le-1.0$ & $400<D\le800$ & $0.640$ & $0.243$ & 0.0043 & $0.429$ & $0.440$ & 0.3963 & $0.211$ & $0.503$ &  0.42 \\
 & $800<D\le1000$ & $0.707$ & $0.303$ & 0.0151 & $0.346$ & $0.386$ & 0.4006 & $0.360$ & $0.491$ &  0.73 \\
 \cline{2-11}
 & $D\le1000$ & $0.609$ & $0.151$ & 0.0000 & $0.326$ & $0.191$ & 0.0973 & $0.283$ & $0.243$ &  1.16 \\
 \hline
 & $D\le400$ & $0.232$ & $0.228$ & 0.3121 & $0.138$ & $0.263$ & 0.6244 & $0.094$ & $0.348$ &  0.27 \\
$-1.0<\Delta M_r\le0.0$ & $400<D\le800$ & $0.443$ & $0.184$ & 0.0183 & $0.622$ & $0.243$ & 0.0059 & $-0.179$ & $0.305$ &  0.59 \\
 & $800<D\le1000$ & $0.895$ & $0.386$ & 0.0027 & $0.547$ & $0.303$ & 0.0813 & $0.348$ & $0.491$ &  0.71 \\
 \cline{2-11}
 & $D\le1000$ & $0.552$ & $0.142$ & 0.0000 & $0.560$ & $0.154$ & 0.0001 & $-0.007$ & $0.209$ &  0.03 \\
 \hline
 & $D\le400$ & $0.530$ & $0.193$ & 0.0031 & $0.566$ & $0.152$ & 0.0001 & $-0.036$ & $0.245$ &  0.15 \\
$0.0<\Delta M_r\le1.0$ & $400<D\le800$ & $0.470$ & $0.150$ & 0.0006 & $0.525$ & $0.170$ & 0.0010 & $-0.055$ & $0.226$ &  0.24 \\
 & $800<D\le1000$ & $0.517$ & $0.243$ & 0.0278 & $0.762$ & $0.191$ & 0.0000 & $-0.244$ & $0.309$ &  0.79 \\
 \cline{2-11}
 & $D\le1000$ & $0.494$ & $0.113$ & 0.0000 & $0.572$ & $0.122$ & 0.0000 & $-0.079$ & $0.167$ &  0.47 \\
\enddata
\tablecomments{$R_C$ is the Pearson correlation coefficient $\pm$ bootstrap uncertainty for coherent neighbors, while $R_{AC}$ is for anti-coherent neighbors. The $p_C$ and $p_{AC}$ indicate the $p$-values for coherent and anti-coherent neighbors, respectively. The last column ($\sigma$) shows the statistical significance of the difference between the $R$ values for coherent and anti-coherent neighbors ($R_C - R_{AC}$).}
\label{diff}
\end{deluxetable*}

Figure~\ref{col9} shows the correlations in $g-r$ color between the CALIFA galaxies and their neighbors in various relative luminosity and projected distance bins. The correlation coefficients ($R$) for coherent and anti-coherent neighbors are respectively estimated in each bin. 
It is noted that the coherent neighbors obviously outnumber the anti-coherent neighbors at $\Delta M_r\le-1.0$ (45 versus 27; 166$\%$), whereas such superiority in number is weak at $-1.0<\Delta M_r\le0.0$ (57 versus 44; 130$\%$) or absent at $0.0<\Delta M_r\le1.0$ (97 versus 106; 92$\%$), which reconfirms that the small-scale dynamical coherence is dominated by bright neighbors.

For bright neighbors ($\Delta M_r\le-1.0$), in Figure~\ref{col9}, the correlation coefficient with coherent neighbors tend to be larger than that with anti-coherent neighbors. At any distance bin within 1 Mpc, the bright and coherent neighbors show moderate or strong correlations ($R\ge0.64$) in $g-r$ color with the CALIFA galaxies with tiny $p$-values ($p\lesssim 0.015$). On the other hand, the bright but anti-coherent neighbors hardly show any meaningful correlation: the $p$-values are very large ($R\le0.43$ and $p\gtrsim0.3$). Meanwhile, for intermediate or faint neighbors ($\Delta M_r>-1.0$), the difference between coherent and ant-coherent neighbors does not appear consistently along the $D$ bins. The consistent difference is visible only for bright neighbors.

When the bins are divided only by the relative luminosity (i.e., merging Figure~\ref{col9} along $D$ out to 1 Mpc), the difference between coherent and anti-coherent neighbors is relatively large (0.609 versus 0.326 in $R$, and 0.00 versus 0.10 in $p$-value) for bright neighbors ($\Delta M_r\le-1.0$), whereas the difference is almost negligible for other cases, as shown in Figure~\ref{col3}. For intermediate or faint neighbors, both of the coherent and anti-coherent neighbors show moderately and similarly positive correlations with the CALIFA galaxies.

Although Figure~\ref{col9} seems to support the idea that the coherent neighbors have stronger conformity with the CALIFA galaxies than the anti-coherent neighbors (when the neighbors are bright), it is necessary to examine how statistically significant such differences are. For this purpose, the $p$-value is not an appropriate indicator, because it does not tell us how easily the measured correlation can be reproduced by a random selection of sample galaxies.
To resolve this, we estimated the bootstrap uncertainty of each correlation coefficient, by randomly sampling the same number of CALIFA - neighbor pairs from the whole sample with replacement, regardless of coherence, distance and relative luminosity. The random sampling was repeated 1000 times, from which the standard deviation of $R$ values is regarded as its statistical uncertainty.

The results are presented in Figure~\ref{rcomp}. In addition, Table~\ref{diff} summarizes how significantly different the correlation coefficients are in each bin of relative luminosity and projected distance. When we consider the bootstrap uncertainties, the difference between coherent and anti-coherent neighbors is significant ($2.6\sigma$) only for bright ($\Delta M_r\le-1.0$) and close ($D\le400$ kpc) neighbors. For the other cases, the difference is statistically insignificant and not even marginal ($<1\sigma$). Therefore, from these results, it can be argued that the dynamical coherence is accompanied with galaxy conformity only when the neighbors are sufficiently bright and close.

\begin{deluxetable}{ccc|ccc}
\tablenum{2} \tablecolumns{6} \tablecaption{$g-r$ colors of the CALIFA galaxies and their neighbors in Figure~\ref{col9}(a)} \tablewidth{0pt}
\tablehead{  \multicolumn{3}{c}{Coherent} & \multicolumn{3}{c}{Anti-coherent} \\ ID & CALIFA & Neighbor & ID & CALIFA & Neighbor }
\startdata
 & 0.7482 & 0.8455  & & 0.5376 & 0.9808 \\
 & 0.7065 & 0.8966  & & 0.3655 & 0.9978 \\
 & 0.5888 & 0.6667  & & 0.6079 & 0.9980 \\
 & 0.7099 & 0.8288  & & 0.6906 & 0.6362 \\
 & 0.2914 & 0.6813  & & 0.7618 & 0.8158 \\
 & 0.4134 & 0.6347  & & 0.7906 & 0.9292 \\
 & 0.8983 & 0.8153  & & 0.3350 & 1.0070 \\
 & 0.6267 & 0.8073  & & 0.5414 & 0.8215 \\
 & 0.4336 & 0.3950  & & & \\
 & 0.6343 & 0.7953  & & & \\
(1) & 0.7245 & 0.7576  & (1) & 0.7245 & 0.7863 \\
(2) & 0.8141 & 0.8294  & (2) & 0.8141 & 0.8288 \\
(3) & 0.7425 & 0.8294  & (3) & 0.7425 & 0.8288 \\
(4) & 0.2844 & 0.6305  & (4) & 0.2844 & 0.7021 \\
(5) & 0.7342 & 0.7037  & (5) & 0.7342 & 0.6871 \\
(5) & 0.7342 & 1.0426  & & & \\
\enddata
\tablecomments{Five CALIFA galaxies have multiple neighbors in the ranges of $D\le400$ kpc and $\Delta M_r \le -1.0$ mag, which are denoted by temporary ID numbers (1 to 5). ID-(5) has three neighbors. }
\label{multi}
\end{deluxetable}

As mentioned in Section~\ref{data}, a single CALIFA galaxy may have multiple neighbors in our sample selection. We checked how significant such duplication is in Figure~\ref{col9}(a). Table~\ref{multi} lists the $g-r$ colors of the CALIFA galaxies and their neighbors in the ranges of $D\le400$ kpc and $\Delta M_r \le -1.0$ mag. The actual number of CALIFA galaxies included in Figure~\ref{col9}(a) is counted to be 23, among which five CALIFA galaxies have multiple neighbors. Interestingly, every case of the five CALIFA galaxies has both of coherent and anti-coherent neighbors. One case (ID-5 in Table~\ref{multi}) has three neighbors: two coherent neighbors and one anti-coherent neighbor. It is noted that ID-(2) and ID-(3) share their two neighbors, and the whole four galaxies seem to have similar colors regardless of their coherence. When all of the duplicated CALIFA galaxies are eliminated from the sample, the correlation coefficients (p-values) are estimated to be 0.686 (0.029) and $-0.536$ (0.171) for coherent and anti-coherent neighbors, respectively: the difference still seem to be obvious.

We visually checked the appearances of the CALIFA galaxies that have bright ($\Delta M_r\le-1.0$), close ($D\le400$ kpc) and coherently moving neighbors (several examples are shown in Figure~\ref{atlas}). Among them, we found only one CALIFA galaxy (K0520) that shows obvious features of ongoing interactions with its bright neighbor that is as close as 16 kpc. In the other CALIFA galaxies, which have bright neighbors at 30 kpc or farther distances, we hardly find obvious tidal features that stretch toward the bright neighbors. It is noted that a couple of galaxies (e.g., K1010, K2052, and K3021) show some distorted internal structures that may indicate their recent tidal events, although it cannot be confirmed here if they originate from the interactions with the bright neighbors (directed by the arrows) or not.

\section{DISCUSSION AND CONCLUSION}\label{discuss}

From our investigation, one striking result is that the coherent neighbors show significantly stronger conformity in $g-r$ color with the CALIFA galaxies than the anti-coherent neighbors, when the neighbors are bright ($\Delta M_r\le-1.0$) and close ($D\le400$ kpc). 
This result is in agreement with the conclusion of \citet{lee19a} that the small-scale dynamical coherence may originate from recent interactions between galaxies, in the context that bright ($\sim$ massive) neighbors at close distances may strongly influence both of the rotations (dynamical coherence) and colors (galaxy conformity) of the CALIFA galaxies.
As reported in previous studies about galaxy conformity \citep[e.g.,][]{par08a}, the projected distance between two galaxies is known to be a key factor to affect the conformity between them. In this paper, we argue that the dynamical coherence is another important factor: even among the neighbors at similarly close (projected) distances, the neighbors moving coherently with the rotational direction of a given galaxy influence the stellar populations of the galaxy more strongly than the anti-coherently moving neighbors do.

Then, why do anti-coherent neighbors show weaker galaxy conformity? A probable answer is that the anti-coherent neighbors may not have actually interacted with the given galaxy yet: their distance in the three-dimensional space may be much farther than the projected distance. Otherwise, the anti-coherent neighbors may have just come close to the given galaxy (the first encounter), while the coherent neighbors may have had sufficient time to interact, possibly because they have already passed by the given galaxy (in the backsplash stage, so called). In any case, the point is that the dynamical coherence may be partial (but not perfect) evidence of a recent interaction with a neighbor that is more massive than the target (CALIFA) galaxy. That is, the dynamical coherence would be a new indicator if two close galaxies have actually interacted or not.

For example, when there are two galaxies that we want to test if they have recently interacted with each other but they do not show visible tidal features between them (like most cases in Figure~\ref{atlas}), it will be helpful to estimate the line-of-sight motion of the brighter galaxy relative to the fainter galaxy, and the rotation of the fainter galaxy. Even if we find dynamical coherence between the two galaxies, this test may not confirm that they have experienced a recent interaction, because the coherence may happen simply by chance. On the other hand, if they do not show dynamical coherence (i.e., unless the rotation of the fainter galaxy is coherent with the line-of-sight motion of the brighter one), major interactions of stellar bodies between them probably did not happen yet.
For such a test, the IFS observations will be the best, but long-slit spectroscopy across the two galaxies can be used as an alternative method with lower cost.

It is noted that anti-coherent neighbors also show some moderate conformity signals in most cases. This may be because the galaxy conformity is not necessarily the result of direct interactions of stellar bodies. According to \citet{par08b}, gas interaction as well as the interaction of stellar bodies can cause conformity between close pair galaxies. Moreover, since faint neighbors are expected to hardly influence the rotation of a given galaxy as reported in \citet{lee19a}, the anti-coherence of a faint neighbor may not be the evidence that the neighbor do not interact with the given CALIFA galaxy. In this regard, on the contrary, faint and coherent neighbors cannot be asserted to be the neighbors that have actually interacted. In other words, the dynamical coherence may be an indicator for the actual interactions with bright neighbors, but not necessarily for the interactions with faint neighbors.

Finally, we discuss some caveats in our analysis. First, our results are based on a galaxy sample with a small size. Although the conformity difference between coherent and anti-coherent neighbors appears to be statistically meaningful for bright and close neighbors (Figure~\ref{col9}(a) and Table~\ref{diff}), the $2.6\sigma$ significance may not be high enough to perfectly confirm our argument that galaxy conformity tends to accompany dynamical coherence. Follow-up studies with a much larger sample, hopefully covering out to $2R_e$ in each IFS target, will be required for higher significance.
Second, there is a fundamental limit in exactly identifying the neighborhood between galaxies and the influence from the neighborhood. Since it is almost impossible to estimate three-dimensional position and velocity of a galaxy in observation, we cannot but depend on some assumptions: the two-dimensional projection is a part of such assumptions. Moreover, to reduce uncertainty, we adopted the X-cut selection of neighbors as described in Section~\ref{data}, but it would exclude a half of actual neighbors of a given CALIFA galaxy. Such projection and selection are necessary in a statistical study and they work to some extent, but it needs to be kept in mind that the excluded neighbors may have some influence on the results, which may not be entirely negligible particularly when the sample size is small as in our work.

In conclusion, the dynamically coherent neighbors show significantly stronger conformity than the dynamically anti-coherent neighbors, when they are sufficiently bright ($\Delta M_r \le-1.0$) and close ($D\le400$ kpc). This result agrees with the previous conclusion of \citet{lee19a}: recent interactions with bright neighbors are the main drivers of the small-scale dynamical coherence.
This finding would be usefully applied to constraining whether two close pair galaxies have actually interacted with each other or not, when their tidal features are hardly visible, by observing them with IFS or long-slit spectroscopy.

\acknowledgments
This work was supported by UST Young Scientist Research Program through the University of Science and  Technology (No.~2019YS10).
This study uses data provided by the Calar Alto Legacy Integral Field Area (CALIFA) survey (http://califa.caha.es/), which is based on observations collected at the Centro Astron{\'o}mico Hispano Alem{\'a}n (CAHA) at Calar Alto, operated jointly by the Max-Planck-Institut f{\"u}r Astronomie and the Instituto de Astrof{\'i}sica de Andaluc{\'i}a (CSIC).

\end{document}